\documentclass{aastex61}

\usepackage{natbib}
\usepackage{soul}
\usepackage{graphicx,float,amsmath,bm,epsfig,epstopdf}

\shortauthors{Pal et al.}

\begin{document}

\title{A Sun-to-Earth analysis of magnetic helicity of the 17-18 March 2013 Interplanetary coronal mass ejection}

\correspondingauthor{Nat Gopalswamy}
\email{nat.gopalswamy@nasa.gov}

\author{Sanchita Pal}
\affiliation{Center of Excellence in Space Sciences India, Indian Institute of Science Education and Research Kolkata\\
Mohanpur 741246, West Bengal, India }

\author{Nat Gopalswamy}
\affiliation{NASA Goddard Space Flight Center  \\
Greenbelt, MD 20771, USA }

\author{Dibyendu Nandy}
\affiliation{Center of Excellence in Space Sciences India, Indian Institute of Science Education and Research Kolkata\\
Mohanpur 741246, West Bengal, India }
\affiliation{Department of Physical Sciences, Indian Institute of Science Education and Research Kolkata\\
Mohanpur 741246, West Bengal, India }

\author{Sachiko Akiyama}
\affiliation{NASA Goddard Space Flight Center  \\
Greenbelt, MD 20771, USA }
\affiliation{The Catholic University of America\\
Washington DC 20064, USA}

\author{Seiji Yashiro}
\affiliation{NASA Goddard Space Flight Center  \\
Greenbelt, MD 20771, USA }
\affiliation{The Catholic University of America\\
Washington DC 20064, USA}

\author{Pertti Makela}
\affiliation{NASA Goddard Space Flight Center  \\
Greenbelt, MD 20771, USA }
\affiliation{The Catholic University of America\\
Washington DC 20064, USA}

\author{Hong Xie}
\affiliation{NASA Goddard Space Flight Center  \\
Greenbelt, MD 20771, USA }
\affiliation{The Catholic University of America\\
Washington DC 20064, USA}

 \begin{abstract}

We compare the magnetic helicity in the 17-18  March 2013 interplanetary coronal mass ejection (ICME) flux-rope at 1 AU and in its solar counterpart. The progenitor coronal mass ejection (CME) erupted on 15 March 2013 from NOAA active region 11692 and associated with an M1.1 flare. We derive the source region reconnection flux using post-eruption arcade (PEA) method \citep{2017aSoPh..292...65G} that uses the photospheric magnetogram and the area under the PEA. The geometrical properties of the near-Sun flux rope is obtained by forward-modeling of white-light CME observations. Combining the geometrical properties and the reconnection flux we extract the magnetic properties of the CME flux rope \citep{2017barXiv170508912G}. We derive the magnetic helicity of the flux rope using its magnetic and geometric properties obtained near the Sun and at 1 AU. We use a constant-$\alpha$ force-free cylindrical flux rope model fit to the in situ observations in order to derive the magnetic and geometric information of the 1-AU ICME. We find a good correspondence in both amplitude and sign of the helicity between the ICME and the CME assuming
a semi-circular (half torus) ICME flux rope with a length of $\pi$ AU. We find that about 83\% of the total flux rope helicity at 1 AU is injected by the magnetic reconnection in the low corona. We discuss the effect of assuming flux rope length in the derived value of the magnetic helicity. This study connecting the helicity of magnetic flux ropes through the Sun-Earth system has important implications for the origin of helicity in the interplanetary medium and the topology of ICME flux ropes at 1 AU and hence their space weather consequences. 
 \end{abstract}

\keywords{Sun: magnetic helicity, Sun: coronal mass ejection, magnetic cloud, solar wind}

\section{Introduction} \label{sec:intro}

Magnetic clouds (MCs) are a special kind of interplanetary manifestation of coronal mass ejections (CMEs) characterised by 1) a low proton temperature and a low proton beta compared to the typical solar wind, 2) an enhanced magnetic field strength and 3) a smoothly rotated magnetic field \citep{1981JGR..86..6673B}. The flux rope magnetic structure of MCs makes them a distinct subset of interplanetary coronal mass ejections (ICMEs). The characteristics of MCs are of great interest because about $90\%$ of the observed MCs are responsible for driving geomagnetic storms with Dst$_{min}$ (minimum Dst index observed in geomagnetic storm interval) $<= -30$ nT \citep[e.g.][]{2007SoPh..242..159W, 2008AdSpR..41..335W}.\par

CMEs are generated due to the destabilization of coronal magnetic field structure, usually triggered by magnetic field evolution due to flux emergence, twisting, shearing and converging motions in the photosphere. Magnetic reconnection in the solar corona allows transferring the photospheric shear to the twisted magnetic flux rope \citep{1996SoPh..167..217L, 2006SoPh..233..3D}. The coronal flux rope thus formed becomes an MC after propagating through the interplanetary medium.  Several studies have shown that all ICMEs have plausibly flux rope structure \citep{2013SoPh..284...17G,2015SoPh..290..1371M}, which is an important fact that can be used for space weather predictions. However only in situ observations of CMEs are not sufficient to conclude that flux rope configuration is present in all CMEs \citep{2011ApJ...738L..18A,2013SoPh..284..179V}. \par
One of the most important characteristics of CMEs is the magnetic helicity, which describes how the magnetic flux tubes are twisted, linked and wounded around each other in a closed volume. In the solar atmosphere and heliosphere the magnetic helicity is almost conserved \citep{1984JFM..147..133B}. Theoretical considerations based on the conservation of magnetic helicity generate important constraints on flux tube dynamics in the solar convection zone and the tilt and twist of solar active regions (ARs) \citep{2006JGRA..11112S01N}. Beyond the solar atmosphere, throughout the propagation of MCs in the interplanetary medium, the magnetic helicity also remains invariant in a closed volume unless they significantly reconnect with the surrounding interplanetary magnetic field (IMF). This approximation has been the basis of several studies that connect interplanetary flux ropes with their solar sources \citep{2005Sol..592..605D, 2007ApJ...659..758Q, 2013SoPh..284..105C, 2014ApJ...793...53H}.\par

The computation of magnetic helicity of a flux rope requires the estimation of its geometry and axial magnetic field intensity \citep{2002A&A...382..650D,2000ApJ...539..944D}. Based on the observed data at 1-AU it is possible to analyze the MC geometry by fitting a cylindrical or a torus-shaped flux rope model. Thus the axial magnetic field strength can be estimated. Considering the geometry of the causative CME and its poloidal flux the helicity of the progenitor coronal flux rope can be measured. Several studies have shown that the low-coronal reconnection flux is almost equivalent to the azimuthal flux of the flux rope formed due to reconnection \citep{2007ApJ...669..621L,2007ApJ...659..758Q,2017aSoPh..292...65G}. \citet{2017aSoPh..292...65G} devised a convenient method to compute the reconnection flux using the area under the post eruption arcade (PEA) and the photospheric flux threading through this area.   \par

The main aim of this study is to compare the magnetic properties of an MC with its associated coronal flux rope. In particular, the magnetic helicity in the two domains is compared. For this study, we choose an identified MC event observed on 17 March 2013 at 1 AU with a distinct solar source having a clear coronal signature and available line-of-sight and vector magnetograms. Next, we extract the geometrical and magnetic information of the MC and its solar counterpart. Finally, we comment on the conservation of magnetic helicity in the Sun-Earth system based on our study and reflect upon the major source of magnetic helicity of CME flux ropes.

\section{ Observations}
\subsection{in situ observation of ICME at at 1 AU}

Figure 1 shows the in situ solar wind plasma and magnetic properties measured by the Advanced Composition Explorer (ACE) spacecraft \citep{1998SSRv...86....1S} \url{(http://www.srl.caltech.edu/ACE/ASC/level2/lvl2DATA_MAG-SWEPAM.html)} at L1 Lagrangian point along with the geomagnetic activity index (Dst) \url{(http://wdc.kugi.kyoto-u.ac.jp/dst_final/index.html)} during 17-18 March 2013. Starting from the top, we plot the total IMF intensity ($B$), Y, Z components of IMF in Geocentric Solar Ecliptic (GSE) coordinate defined as $B_{y}$ and $B_{z}$, solar wind plasma flow speed ($V_{sw}$), density ($N_p$), proton temperature ($T_{p}$), and plasma beta,  computed based on protons, i.e. proton beta (beta) with a 64-second time resolution and Dst with 1 hour time resolution. The dashed curves over-plotted on $T_{p}$ represent the temperature ($T_{ex}$) expected from the correlation between $V_{sw}$ and $T_p$ \citep{1987JGR....92..11189L}. The horizontal dashed line over-plotted on beta stresses the distinct value, beta=1. A sudden enhancement of $V_{sw}$, $N_p$, $T_{p}$ and $B$ implies that an ICME arrives with an interplanetary (IP) shock at 05:30 UT on 17 March 2013 (marked by a dotted vertical line in Figure 1.) The observed shock velocity is 751 km/s. A sheath region with a high fluctuation of IMF vectors is present after the IP shock. The peak value of the IMF intensity and the minimum value of its z component ($B_{z}$) in this region are respectively 22.1 nT and -18.5 nT. After the sheath region, a decrease in $T_{p}$ compared to $T_{ex}$, beta with value less than 1, constant westward (negative) $B_{y}$ and south to north (negative to positive) smooth rotation of $B_{z}$ with strong magnetic field strength confirms that a small inclination bipolar (south-north) magnetic cloud \citep{1990JGR....95..17267G,2004JASTP..66..323L} passes through L1. Moreover, the decrease in $T_{p}$ compared to $T_{ex}$ is only present close to the beginning and at the end of the MC. Within the identified boundaries, $T_p$ is not below $T_{ex}$ during some interval, although beta is $<1$.  However, the speed has the declining profile indicative of MC expansion; the By component shows the MC axis points to the west, while the Bz component shows rotation from south to north. Consistent with the \citet{1981JGR..86..6673B} definition the MC starts at 14:39 UT (marked by the red vertical line) and ends at 00:44 UT on 18 March (marked by the blue vertical line). We note that the rear boundary is well defined in this case, indicated by the discontinuities in the total magnetic field and solar wind speed. However, it must be noted that frequently the MC boundaries are not well defined. The duration of the MC is about 10 hr, which is $\approx48\%$ smaller than the average duration of solar cycle 24 MC at 1 AU, $\Delta t_{mc}$= 19.19 hr \citep{2015JGRA..120.9221G}. The plasma speed at the MC's leading edge and its trailing edge are respectively 671 and 567 km/s \citep{2015JGRA..120.9221G}. From these values, we derive the cloud's central speed as 619 km/s and the expansion speed as 104 km/s. The peak field strength and the minimum value of $B_{z}$ during MC interval are recorded as 13.4 nT and -10.6 nT, respectively. The ICME resulted in a classic double-dip, major geomagnetic  storm \citep{1998JGR...103.6917K} due to southward IMF in the sheath, and MC. The minimum value of Dst is -100 nT during the sheath and -134 nT during the MC.

\begin{figure}
\centering
  \includegraphics[width=0.7\textwidth]{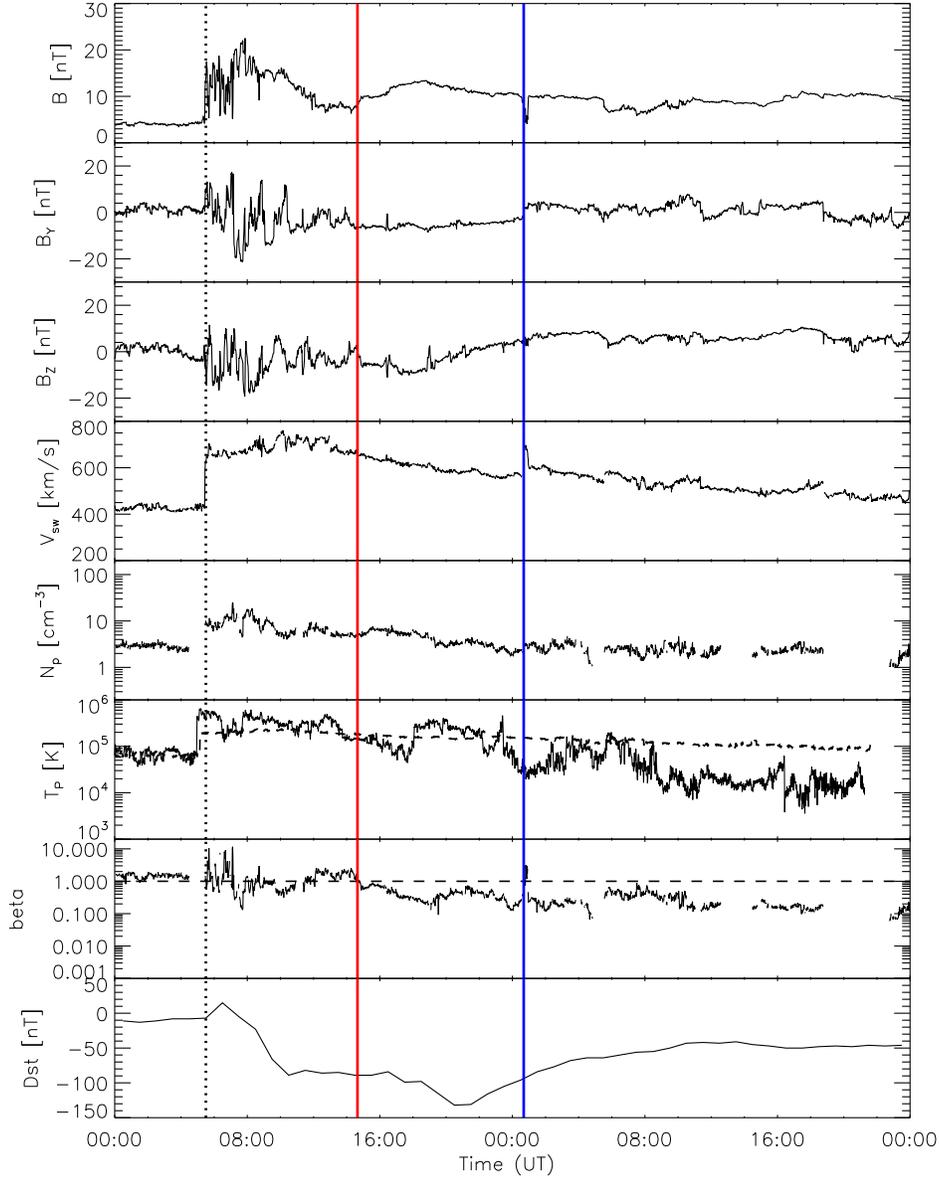}
   \caption{Solar wind data along with geomagnetic storm index (Dst) during 17-18 March 2013. The parameters plotted (from top to bottom) are the total IMF intensity ($B$ in nT), y component of IMF ($B_y$ in nT), z component of IMF ($B_z$ in nT), solar wind plasma flow speed ($V_{sw}$ in km/s), proton density ($N_p$ in cm$^{-3}$), proton temperature ($T_p$ in K), proton beta (beta) and Dst index in nT. The dashed curve over-plotted on $T_p$ represents $T_{ex}$ and the horizontal dashed line on beta points the value, beta=1. The black (dotted), red and blue vertical lines indicate the shock arrival, start and the end boundary of MC, respectively.}
\end{figure}

\subsection{Identification and observation of ICME solar source}

To identify the solar source of the ICME, we use observations from different instruments on the Solar and Heliospheric Observatory (SOHO) and Solar Dynamic Observatory (SDO) missions. The Large Angle and Spectrometric Coronagraph (LASCO) telescope's C2 and C3 on board SOHO observe the CME near the Sun. The fields of view of C2 and C3 \citep{1998GeoRL..25..3019B} are respectively $2-6$ and $4-30$ R$_s$ measured in units of solar radius from the disk center of the Sun. We use the LASCO CME catalog \url{(http://cdaw.gsfc.nasa.gov/CME_list/)} \citep{2004JGRA..109.7105Y,2009EM&P..104..295G} to identify the MC-associated CME. To analyze the structure of the identified source CME, we use the observations from the Sun Earth Connection Coronal and Heliospheric Investigation \citep[SECCHI,][]{2008SSRv..136...67H} COR2 A \& B on board the Solar Terrestrial Relations Observatory (STEREO) mission along with the LASCO images. We study the solar source of MC-associated CME by using SDO's Atmospheric Imaging Assembly (AIA) \citep{2012SoPh..275...17L} images at 193 \AA,  Helioseismic Magnetic Imagers (HMI) \citep{2012SoPh..275..207S} line-of-sight (LOS) magnetogram and the Space-Weather HMI AR Patch (SHARP) \citep{2014SoPh..289.3549B} vector magnetogram to study the source active region of the ICME. \par

\begin{figure}
\centering
 \includegraphics[width=0.7\textwidth]{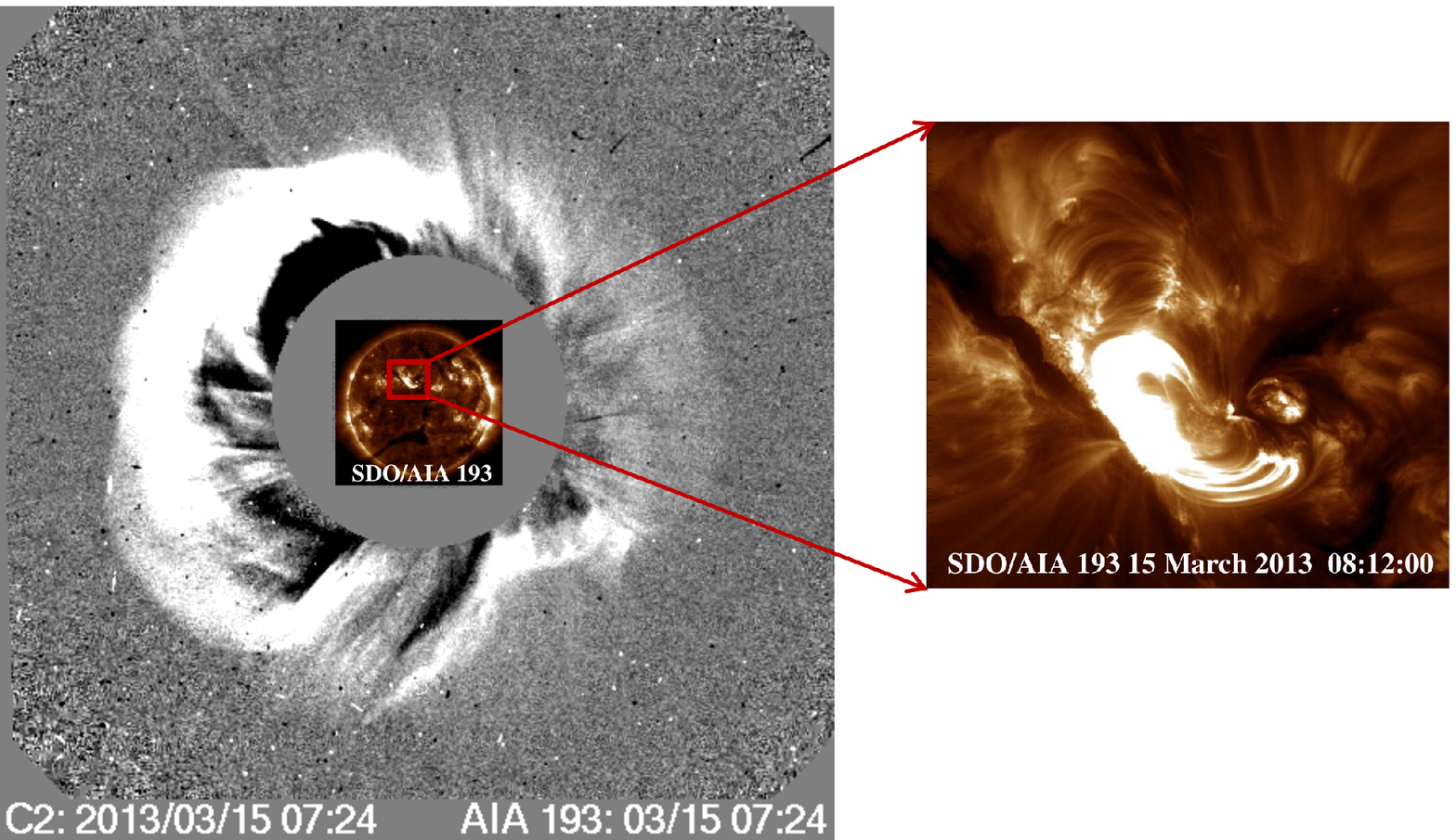}
   \caption{The 15 March 2013 CME in LASCO/C2 difference image (left) with PEA as a coronal signature shown in an SDO/AIA image at 193 \AA\ (right).}
\end{figure}

We identify the  CME associated with the ICME in question by following the procedure described in \citet{2007JGRA..11212103Z}. We consider the solar wind speed at shock arrival (757 km/s) as the transit speed of the CME-driven shock travels from the Sun to the Earth and estimate the transit time as $\approx54$ hrs, suggesting that the associated CME should start only after 00:00 UT of 15 March 2013. By searching the LASCO CME catalog for frontside wide CMEs during the transit time interval we find only one CME originating close to the disk center in the LASCO/C2 FOV at 07:12 UT on 15 March 2013. It has a high projected speed of 1063 km/s consistent with the fast ICME. The CME is associated with an M1.1 GOES soft X-ray flare that initiated at 05:46 UT on 15 March. The CME leaves behind a PEA as an apparent coronal feature observed by SDO/AIA at 193 \AA\ during the decay phase of the long duration flare. Thus we identify the source location of the CME as N11E12 which corresponds to the NOAA AR 11692. In Figure 2 we show the running difference image of the associated CME observed by LASCO/C2 along with the PEA in solar corona observed by SDO/AIA at 193 \AA.

\section{Analysis and Results}

\subsection{Analysis of ICME data}
  
To understand the structure and magnetic nature of the cloud, we use a constant-$\alpha$ (linear) force-free cylindrical flux rope model with self-similar expansion, which was proposed by \citet{1992sws..coll..611F,1993JGR....98.7621F} with a modification based on \citet{2002EP&S...54..783S}. Following \citet{2007AG..25..2453M}, we perform the fitting of the model during the interval, when the magnetic field rotates smoothly in the Y-Z plane, proton temperature and proton beta decrease from their average values, $He^{++}/H^{+}$ value increases, proton number density decreases, and the fluctuation in the ratio of standard deviation to the mean magnetic field intensity ($S_b/B$) is relatively small. Figure 3 shows the solar wind data during 17-18 March 2013 with the cylindrical model fitting results. From top to bottom we plot $B$, $B_x$, $B_y$, $B_z$, $S_b/B$, $V_{sw}$, $N_p$, $T_{p}$, beta, and vector plots of the magnetic field projected on X-Y, X-Z, and Y-Z plane in 30-min average. The vertical solid lines in the figure denote the start and end times of the cloud. The red curves over-plotted on $B$, $B_x$, $B_y$, $B_z$ and $V_{sw}$ represent the fit. $T_{ex}$ is over-plotted on $T_{p}$ in a dashed curve. Table 1 lists the best fit parameters such as, latitude ($\theta_{mc}$) and longitude angles ($\psi_{mc}$) of the axial magnetic field, axial field strength ($B_{0mc}$), radius ($R_{0mc}$) of the fitted cylinder, handedness of the twisted field ($D$) of the cloud, and impact parameter ($p$) in column 1-6. The impact parameter is the distance between the spacecraft trajectory and the MC axis normalized to the MC radius. Column 7 shows the relative error of the fitting ($E_{rms}$). $E_{rms}$ is the ratio of $\Delta$ and the maximum observed magnetic field intensity, $B_{max}$. Here $\Delta$ is the rms deviation between the observed magnetic field, $B(t_i)$, and the model magnetic field, $B^M (t_i)\ (i= 1, . . . , N)$. \par
\begin{figure}
\centering
  \includegraphics[width=0.7\textwidth]{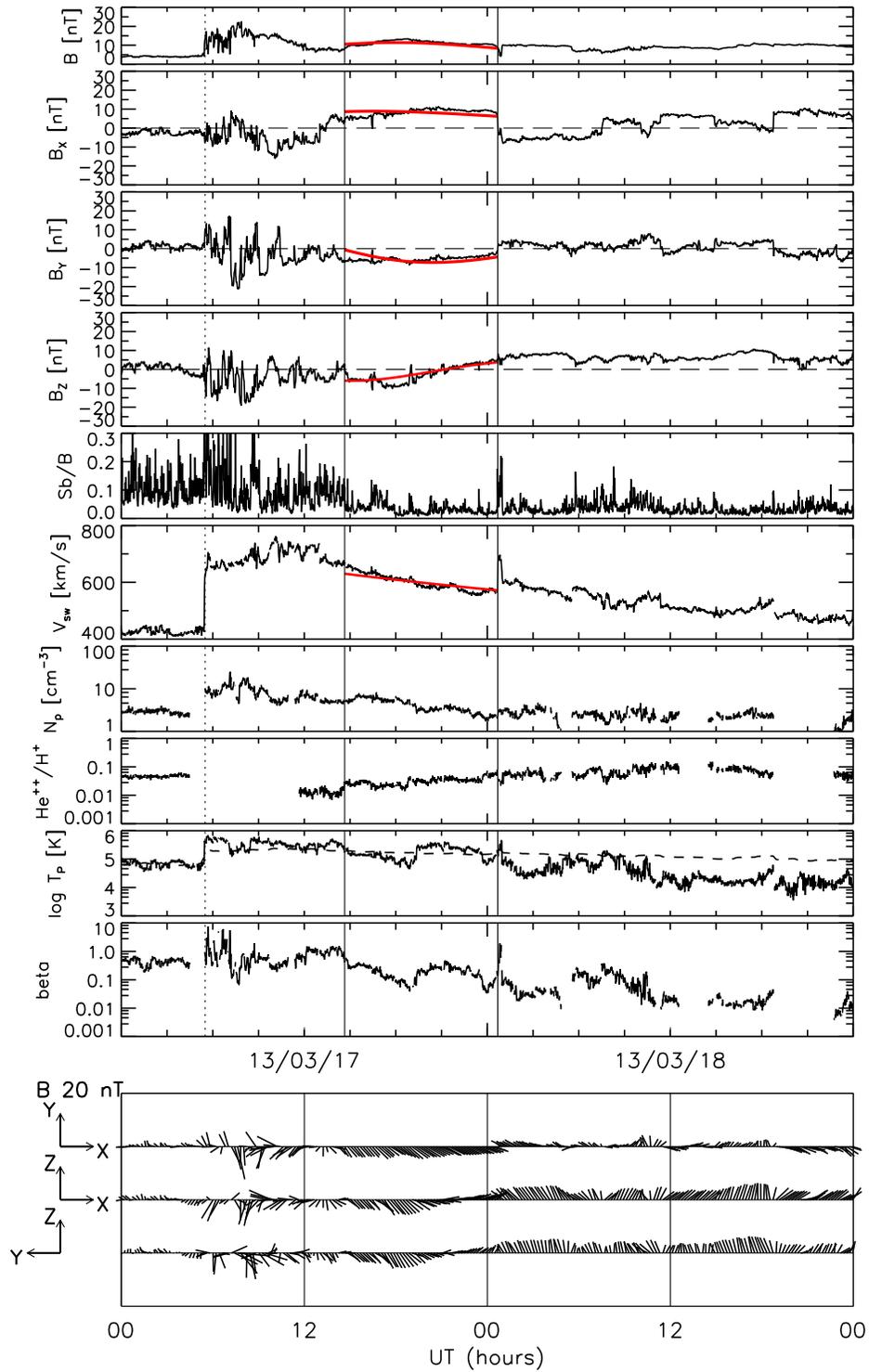}
   \caption{Results of the fitting with cylindrical flux rope model (red curve superimposed on the observed solar wind parameters) to the MC. The continuous vertical lines show the front and end boundary of MC. The vertical dotted line indicates the time of shock arrival. The magnetic field vectors projected on the X-Y, X-Z, and Y-Z planes are shown at the bottom.}
\end{figure}
Figure 4 depicts the geometry of the MC at the time of encounter with spacecraft. In this figure arrow A indicates the axis of the MC flux rope, arrow S shows the direction
of the poloidal magnetic field and arrow denoted by S/C demonstrates the path of spacecraft. It is observed from the figure that the spacecraft comes across the MC near its flank and far below from its axis.

\begin{figure}
\centering
  \includegraphics[width=0.4\textwidth]{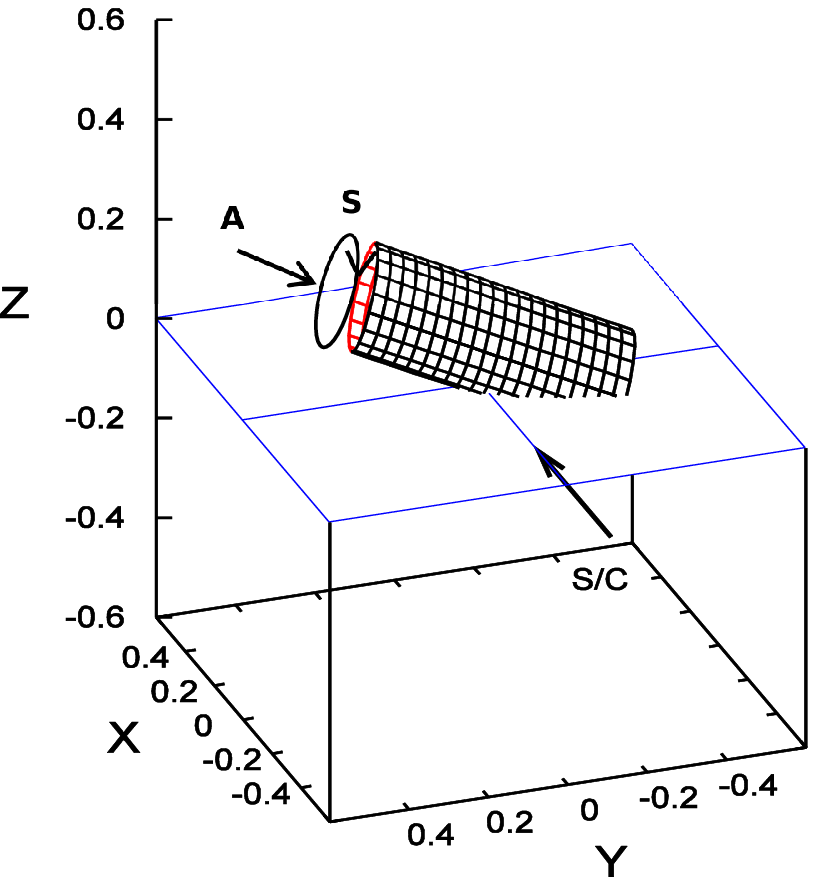}
   \caption{Geometry of the 17-18 March 2013 MC obtained from fitting data. Three directions indicated in figure are axial field direction (A), direction of the poloidal magnetic field (S) and spacecraft trajectory ($S/C$). }
\end{figure}

\subsubsection{Calculation of cloud's magnetic parameters}

    In this section we compute MC unsigned flux and magnetic helicity ($H_{mc}$) of the MC using Lundquist's constant-$\alpha$ force-free field solution in cylindrical coordinates \citep{1990JGR....9511957L}. The solution provides axial magnetic field component, $B_z= B_0 J_0(\alpha r)$, poloidal (azimuthal) magnetic field component, $B_{\theta}= DB_0J_1(\alpha r)$ and radial field component, $B_r= 0$. Where $B_0$ is the axial magnetic field strength, $D= \pm1$ is the flux rope handedness (plus for right handed and minus for left handed), $J_n$ is the $n^{th}$ order Bessel function and $\alpha= x_{01}/R_{0mc}$ is the twist per unit length, where $x_{01}= 2.4048$  is the location of the first zero of $J_0$.\par

The magnetic flux is defined as
\begin{equation}
\phi=\int{ \bm{B} \ dA}
\end{equation}
using cylindrical symmetry, $dA= 2\pi r dr$.

The axial and poloidal components of magnetic flux ($\phi_{z}$ \& $\phi_{p}$) in a cylindrical flux rope is given by \citep[e.g][]{2004JGRA..109.5106L, 2007ApJ...659..758Q},
\begin{equation}
\phi_{z}= 2\pi \int_{0}^{R_{0}}B_{z}\ r dr= \frac{2\pi J_1(x_{01})}{x_{01}} B_{0}R_{0}^2
\end{equation}
and
\begin{equation}
\phi_{p}= 2\pi \int_{0}^{R_{0}}B_\theta\ r dr = \frac{L}{x_{01}}B_{0}R_{0},
\end{equation}
where $R_0$, $B_0$ and $L$ is respectively the radius, axial magnetic field strength and length of cylindrical flux rope.\par
Within a volume $V$ the magnetic helicity $H$ of a field $\bm{B}$ is defined by,
\begin{equation}
H= \int_V \bm{A\cdot B}\ dV,
\end{equation}
where $\bm{A}$ is the vector potential. This definition of helicity is meaningful only if the normal component of $\bm{B}$ ($B_n$) at any surface $S$ surrounding the volume $V$ is zero, i.e. $B_n= 0$. In case of $B_n \neq 0$, relative magnetic helicity ($H_r$) is derived by subtracting the reference magnetic field ($\bm{B}_{ref}$) helicity from $H$ \citep{1984JFM..147..133B}. Thus, $H_r$ is defined by,
\begin{equation}
H_r= \int_{V} \bm{A\cdot B}\ dV -\int_{V} \bm{A}_{ref}\cdot \bm{B}_{ref}\ dV
\end{equation}
Here $H_r$ is gauge-invariant if $\bm{A} \times \hat{n}= \bm{A}_{ref} \times \hat{n}$ on $S$ of $V$. For a cylindrical flux tube, $\bm{B}_{ref}$ can be chosen as $\bm{B}_{ref}= B_z\hat{u_z}$ and $\bm{B}= B_\theta \hat{u_{\phi}}+B_z\hat{u_z}$. Considering $\bm{A}= \bm{B}/\alpha$, we compute the magnetic helicity of cylindrical flux rope \citep{2003and..book..345B,2003JGRA..108.1362D,2002A&A...382..650D,2000ApJ...539..944D} as,
\begin{equation}
H= 4\pi L \int_0^{R_{0}} A_{\theta}B_{\theta}\ rdr= \frac{4\pi B_0^2 L}{\alpha}\int_0^{R_{0}} J_1^2(\alpha r)\ r dr\approx 0.7\ B_{0}^{2}R_{0}^{3}L
\end{equation}
To derive the axial and poloidal components of MC flux ($\phi_{zmc}$ and $\phi_{pmc}$) and the magnetic helicity ($H_{mc}$) of the MC we apply $B_0= B_{0mc}$, $R_0= R_{0mc}$ and $L= L_{mc}$ in equation (2), (3) and (6). Here $L_{mc}$ is the estimated length of the MC.\par

The largest uncertainty in calculating the flux and helicity of an MC arises from the MC flux rope length approximation. \citet{1997GeoRL..24.1911L} estimated the length of MC as 2.5 AU by measuring the travel time of suprathermal electrons moving along with the twisted magnetic field lines. The presence of bidirectional electrons in MCs observed at 1 AU suggests the possibility of MCs rooted on the Sun when they reach at 1 AU \citep{2000JGR...10527261S}. We consider the MC a half torus that has a circular cross section as an approximation to the expanded flux rope, extending from Sun to Earth. Then the length of MC at 1 AU is $\pi$ AU if the major axis length of the torus is 1 AU. We note that the length is only 21\% higher than the statistical value (2.6 $\pm$ 0.3 AU) reported in \citet{2016SoPh..291..531D}. In our study, we measure the cloud's flux and helicity using each of the approximated MC flux rope lengths. In column 8 to 11 of Table 1, we show the twist density ($\alpha_{mc}$) of the magnetic field in MC flux rope, $\phi_{zmc}$, $\phi_{pmc}$ and $H_{mc}$ corresponding to different $L_{mc}$s. \par

\begin {table}
\begin{center}
\caption {MC fit parameters and derived quantities. }
\centering

\begin{tabular}{ p{0.8cm} p{0.8cm} p{0.8cm} p{0.8cm} p{0.4cm}p{0.8cm} p{0.8cm} p{1cm} p{0.6cm} p{0.9cm} p{0.8cm} p{1cm}p{0.2cm} p{0.7cm} p{0.9cm} p{1cm}}
 \hline

 $\theta_{MC}$ &$\psi_{MC}$& $B_{0mc}$ &$R_{0mc}$ &$D$ &$p$&$E_{rms}$ &$\alpha_{mc}$&$\phi_{zmc}$ &&$\phi_{pmc}$ && &&$H_{mc}$& \\

  $(^{\circ})$ &$(^{\circ})$ & (nT) &(AU) & &($R_{0mc}$) & &(Gm$^{-1}$)& $(10^{21}$ Mx) &&$(10^{21}$ Mx )&&& &$ (10^{42}$ Mx$^2$)&\\
  (1)&(2)&(3)&(4)&(5)&(6)&(7)&(8)&(9)&&(10)&&&&(11)&\\

  \end{tabular}
  \begin{tabular}{ p{0.8cm} p{0.8cm} p{0.8cm} p{0.8cm} p{0.4cm}p{0.8cm} p{0.8cm} p{1cm} p{0.6cm} p{0.9cm} p{0.8cm} p{1cm}p{0.2cm} p{0.7cm} p{0.9cm} p{1cm}}
 & & & & && & & & $\phi_{pmc}^{L2(b)}$& $\phi_{pmc}^{L2.5(c)}$&$\phi_{pmc}^{L\pi(d)}$&&$H_{mc}^{L2(e)}$&$H_{mc}^{L2.5(f)}$&$H_{mc}^{L\pi(g)}$\\\cline{10-12}\cline{14-16}
-24.4&247.8&20.30&0.1152&$R^{(a)}$&0.87&0.23&0.14&0.82&4.37&5.46&6.86&&4.47&5.58&7.01\\\hline

\end{tabular}

\end{center}
$^{(a)}$\ '$R$' stands for Right handed rotation of magnetic field.\\
$^{(b)}\ \phi_{pmc}$ derived using $L_{mc}= 2$ AU\\
$^{(c)}\ \phi_{pmc}$ derived using $L_{mc}= 2.5$ AU\\
$^{(d)}\ \phi_{pmc}$ derived using $L_{mc}= \pi$ AU\\
$^{(e)}\ H_{mc}$ derived using $L_{mc}= 2$ AU\\
$^{(f)}\ H_{pmc}$ derived using $L_{mc}= 2.5$ AU\\
$^{(g)}\ H_{mc}$ derived using $L_{mc}= \pi$ AU\\

\end{table}

\subsection{Analysis of the solar source}
In this subsection, we obtain the geometrical and magnetic properties of the flux rope near the Sun to compute the helicity. The geometrical properties are obtained by forward modelling white-light CMEs using the GCS model. The magnetic properties are obtained by equating the reconnected flux in the eruption region to the poloidal flux of the erupted flux rope.

 \subsubsection{The associated CME}

The CME is a halo CME observed from LASCO/C2 coronagraph at 05:12 UT on 15 March 2013. It appears as a flux rope like structure at 08:08 UT in the field of view of SECCHI/COR2 A \& B. \citet{2017barXiv170508912G} showed that the magnetic properties of the coronal flux rope can be obtained by combining the reconnection flux and the geometric properties obtained from flux rope fits to white-light data. \citet{2017barXiv170508912G} used the \citet{2006ApJ...652.1740K} flux rope model to obtain the geometrical properties of coronal flux ropes. Here we fit the CME with the forward modeling technique, Graduated Cylindrical Shell (GCS) model developed by \citet{2006ApJ...652..763T} because three views are available from STEREO and SOHO. Thus we get the half angular width ($\gamma$), aspect ratio ($\kappa$), height of the leading edge ($h$), tilt angle ($\lambda$) with respect to the equator and the propagation longitude ($\psi$) and latitude ($\theta$) of the CME. From these results, we calculate the radius ($R_{0cme}$) of the CME's circular annulus at its leading edge point using the relation, $R_{0cme}/h= 1/(1+(1/\kappa))$ derived using equation (1) in \citet{2006ApJ...652..763T} and 3-D speed from the time evolution of CME leading edge. In Figure 5, we show the GCS model in green wire frame over-plotted on the white-light CME, observed by LASCO/C3, and SECCHI/COR2 A \& B around the same time.\par
We estimate the axial magnetic field strength ($B_{0cme}$) of the CME from its poloidal flux component ($\phi_{pcme}$) by taking $\phi_{p}= \phi_{pcme}$, $L= L_{cme}$ (length of CME flux rope from the photosphere) and $R_0= R_{0cme}$ in equation (3). Thus, $B_{0cme}$ can be defined as,

\begin{equation}
B_{0cme}= \frac{\phi_{pcme}x_{01}}{L_{cme}R_{0cme}}
\end{equation}

Here $L_{cme}$ is estimated from GCS model and can be computed as, $L_{cme}= 2h_{leg}+y(h-h_{leg}/cos\gamma)/2-2R_{\odot}$, where $h_{leg}$ is the height of the legs of CME flux rope computed using equation (3) in \citet{2006ApJ...652..763T}, ($(h-h_{leg}/cos\gamma)/2$) is the radius of the arc of the flux rope, $y=2(\pi/2+\gamma)$ is the arc angle in radians and $R_{\odot}$ represents the solar radius. \citet{2007SoPh..244...45L} demonstrated that  $\phi_{pcme}$ is approximately equal to the magnetic reconnection flux ($\phi_{RC}$) in the low corona. \citet{2017aSoPh..292...65G} proposed a process to estimate the reconnection flux considering the half of the unsigned photospheric flux underlying the area occupied by the PEA (as discussed in the next section).\par
In equation (6) we use $B_0= B_{0cme}$, $R_0= R_{0cme}$ and $L= L_{cme}$ to obtain the magnetic helicity ($H_{cme}$) of the CME flux rope.
In Table 2 we present the GCS fitting results along with $R_{0cme}$ at 10 R$_s$ ($R_{0cme}^{10R_s}$), $L_{cme}$ and 3-D speed ($V_{cme}^{3D}$) of the CME estimated from the fitting.
  \begin{figure}
  \centering
  \includegraphics[width=0.7\textwidth]{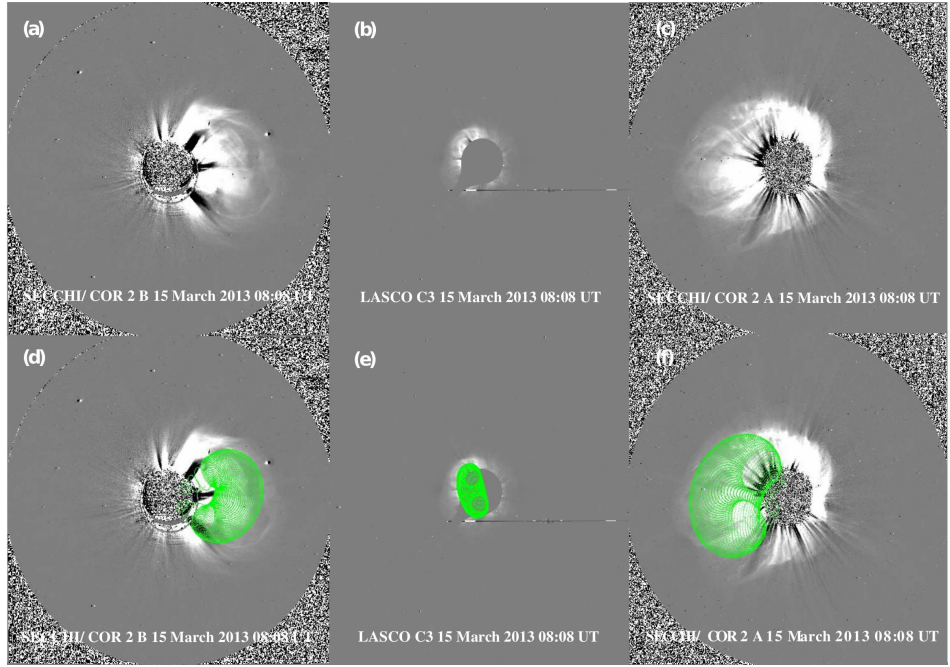}
  \caption{GCS model fitting on 15 March 2013 CME in LASCO/C3, SECCHI A \& B data. (a), (b) and (c) are observed images of the CME by STEREO and LASCO. (d), (e) and (f) show the CME with the GCS wire frame overlaid on it. }
\end{figure}

\begin {table}
\begin{center}
\caption {CME parameters determined by GCS model fitting.}
 \begin{tabular}{ p{6cm} p{1.5cm}  }
 \hline

 \textbf{CME properties} &\textbf{Values} \\
 (1)&(2)\\
 \hline
  $\psi_{cme}$ (Carrington coordinate)     & $68.82 ^{\circ}$ \\
 $\theta_{cme}$ (Carrington coordinate)  & $-6.15 ^{\circ}$ \\
  $\gamma$  & $25.16^{\circ}$\\
 $\lambda$  & $-74.35^{\circ}$\\
 $\kappa$    & 0.27\\

 $R_{0cme}^{10R_s}$ & 2.13 R$_s$\\
 $L_{cme}$ & 17.07 R$_s$\\
 $V_{cme}^{3D}$& 1321 km/s\\

 \hline

\end{tabular}

\end{center}
\end{table}

 \subsubsection{The photospheric source active region}

 To analyze the source active region (AR 11692) of the ICME we consider the SDO/HMI LOS photospheric magnetogram together with the SDO/AIA 193 \AA\ image and the solar X-ray imager aboard the Geostationary Operational Environmental Satellite (GOES) system \citep{2005SoPh..226..255H}, which spatially maps soft X-ray emission of solar corona. We calculate the reconnection flux at the decay phase of flare (when the post eruption arcade is almost matured) by using SDO/HMI LOS, AIA 193 \AA\ data and GOES SXI data. We identify the post eruption arcade (PEA) region in both AIA 193\ \AA\ and GOES SXI images, find the pixels associated with the arcade area and overlay it on HMI LOS data. Considering $B_{LOS}$ and the area of each of those pixels from HMI LOS data we derive the total reconnection flux ($\phi_{RC}$) using the equation \citep{2017aSoPh..292...65G},

 \begin{equation}
\phi_{RC}= \frac{1}{2}\int_{PEA} \mid B_{LOS}\mid dA
\end{equation}

\begin{figure}[htb!]
\centering
  \includegraphics[width=0.7\textwidth]{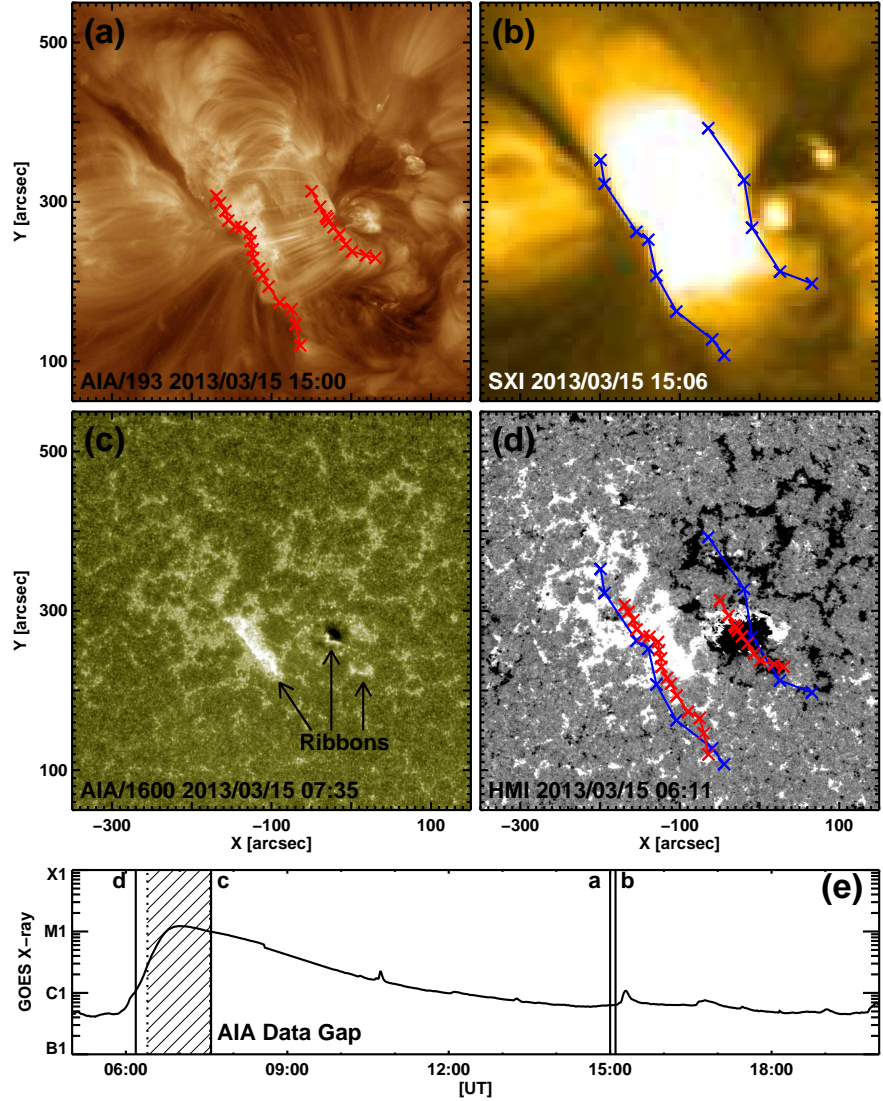}
   \caption{The post eruption arcade, flare ribbon, source magnetogram and flare evolution during 15 March 2013 eruption. (a) SDO/AIA 193 \AA \ image of low corona showing post eruption arcade foot points in red. (b) The PEA observed in GOES SXI and identified foot points in blue. (c) Flare ribbon structure in SDO/AIA/ 1600\AA\ image. (d) HMI LOS photospheric magnetogram with overlaid arcade foot points derived from both AIA and SXI data,  and (e) GOES soft x-ray curve in the 1-8 \AA\ band showing temporal evolution of the M1.1 flare associated with the eruption. The vertical lines marked by a, b, c and d are denoting the epochs when AIA 193\ \AA, SXI, AIA 1600\ \AA\ and HMI LOS data are taken. The shaded region between the dotted line and the vertical line, c defines SDO data gap. $\phi_{RC}$ associated with AIA arcade is $4.10\times10^{21}$ and with SXI arcade is $6.55\times10^{21}$ Mx.}
\end{figure}
To estimate the uncertainty in the arcade area measurement, we consider a range of arcade area ($A_{PEA}$). We derive the upper and lower limits of the range by selecting the arcade foot points in GOES SXI and AIA 193\ \AA\ images. The X-ray imager has a broader temperature response than the EUV image at a particular wavelength.
 In Figure 6(a) and (b), we show the estimated arcade foot points in red and blue lines superposed on AIA 193 \AA\ and GOES SXI images, respectively around 15:00 UT. In Figure 6(d) we overlay the foot points derived from AIA 193\ \AA\ and GOES SXI images on the differential-rotation corrected HMI LOS magnetogram in their respective colors at 6:11 UT (time of available HMI data just at the begining of the flare).  In Figure 6(c) we show the ribbon structures by arrows in AIA 1600\ \AA\ image around 7:35 UT. Figure 6(e) shows the temporal evolution of GOES X-ray flare intensity. The solid vertical lines a, b, and d on the plot indicate the epochs of measuring arcade in AIA 193\ \AA, SXI, and HMI LOS magnetogram. The vertical line c indicates the time when a faded ribbon structure is shown in AIA 1600\ \AA\ image. The shaded interval between the dotted vertical line and the line, c shows an SDO data gap. Due to the absence of SDO/AIA observations during the impulsive phase of the co-produced flare, it is not possible to analyze $\phi_{RC}$ using flare ribbon method as in \citet{2017SPD....4810824K}. In principle, one expects an overestimate of the area of the PEA (and hence $\phi_{RC}$) if the ribbons start at a finite distance from the polarity inversion line (PIL). However, \citet{2017aSoPh..292...65G} considered this issue using events that had both ribbon and PEA data and found no evidence of overestimate. Another possible uncertainty in $\phi_{RC}$ is in identifying the boundaries of PEAs: the PEA is identified in the corona, but superposed on the photosphere. This was also considered by \citet{2017aSoPh..292...65G} and found that the difference is not significant because ribbon and arcade methods give the same value. However, there may be other uncertainties when the arcade appears differently in EUV and X-ray images. This is explicitly shown in Fig. 6: the AIA arcade has a smaller area than the SXI area giving a 37\% lower $\phi_{RC}$. \par

 Using the reconnection flux limits, we derive a range of $B_{0cme}$ at 10 $R_S$ ($B_{0cme}^{10R_s}$) and $H_{cme}$. In Table 3, we show the upper and lower limits of $A_{PEA}$ ($A_{PEA}^u$ \& $A_{PEA}^l$), $\phi_{RC}$ ($\phi_{RC}^u$ \& $\phi_{RC}^l$), $B_{0cme}^{10R_s}$ ($B_{0cme}^{10R_su}$ \& $B_{0cme}^{10R_sl}$, and $H_{cme}$ ($H_{cme}^u$ \& $H_{cme}^l$).\par

\begin {table}
\begin{center}
\caption {Magnetic parameters of 15 March 2013 CME and its source AR derived using HMI LOS, AIA 193 \AA\ and GOES SXI data. }

 \begin{tabular}{ p{1cm} p{1cm}p{1cm} p{1cm} p{1cm}p{1cm}p{1cm} p{1cm}p{1cm} p{1cm} p{1cm}}
  \hline

  \multicolumn{2}{c}{$A_{PEA}$} &&
  \multicolumn{2}{c}{$\phi_{RC}$} &&
  \multicolumn{2}{c}{$B_{0cme}^{10R_s}$}& &
  \multicolumn{2}{c}{$H_{cme}$}\\

  \multicolumn{2}{c}{$(10^{19}$ cm$^2$)} &&
  \multicolumn{2}{c}{$(10^{21}$ Mx)} &&
  \multicolumn{2}{c}{(mG)}& &
  \multicolumn{2}{c}{$(10^{42}$ Mx$^2$)}\\

 \multicolumn{2}{c}{(1)} &&
  \multicolumn{2}{c}{(2)} &&
  \multicolumn{2}{c}{(3)}& &
  \multicolumn{2}{c}{(4)}\\

$A_{PEA}^l$ & $A_{PEA}^u$&& $\phi_{RC}^l$&$\phi_{RC}^u$&&$B_{0cme}^{10R_sl}$ & $B_{0cme}^{10R_su}$& &$H_{cme}^l$ & $H_{cme}^u$\\\cline{1-2}\cline{4-5}\cline{7-8}\cline{10-11}
  9.63 & 20.0 && 4.1& 6.55&& 55.3& 88.4&& 8.5&21.6\\
 \hline

\end{tabular}

\end{center}

\end{table}
In Table 4, we show the helicity difference between the near-Sun and 1-AU flux rope for $L_{mc}$= 2, 2.5, and $\pi$ AU. For each approximated MC length, $H_{mc}$ is less than $H_{cme}$ which suggests that the helicity of the flux rope at 1 AU is primarily invoked by magnetic reconnection at low corona during the eruption. We find a minimum helicity difference in case of $L_{mc}= \pi$ AU with $H_{cme}= H_{cme}^l$. So, the lower boundary of source region helicity is consistent with $H_{mc}$ calculated using the MC length $\approx3.14$ AU. We note that we have used $L_{mc} = 2.5$ AU, which is very close to the average length reported in \citet{2016SoPh..291..531D}. If we use $L_{mc}= 2.6$ AU, the helicity value changes only by a small amount (4\%). We notice a large helicity difference between $H_{cme}^u$ and $H_{mc}$ for each of the $L_{mc}$s. \par

\begin {table}
\begin{center}
\caption {Comparison of magnetic helicity of CME and its interplanetary flux rope counterpart at 1 AU.}
\centering
 \begin{tabular}{ p{2cm} p{2cm} p{2cm}p{0.5cm} p{2cm} p{2cm} p{2cm} }
  \hline
  \multicolumn{3}{c}{$(H_{cme}^l-H_{mc})/H_{cme}^l(\%)$} &&\multicolumn{3}{c}{$(H_{cme}^u-H_{mc})/H_{cme}^u(\%)$}  \\
  \multicolumn{3}{c}{(1)}&&
  \multicolumn{3}{c}{(2)}\\
  \cline{1-3}\cline{5-7}

  $L_{mc}$=2 AU & $L_{mc}$=2.5 AU & $L_{mc}=\pi$ AU & &  $L_{mc}$=2 AU & $L_{mc}$=2.5 AU & $L_{mc}=\pi$ AU  \\\cline{1-3}\cline{5-7}
  47.4&34.4& 17.5&&79.3&74.2 &67.5\\
 \hline
\end{tabular}
\end{center}
\end{table}

\section{Discussion}

Along with the study of magnetic helicity, we measure other nonpotential parameters such as, total unsigned vertical current ($I_z$), mean of global twist density ($\alpha_{AR}$) \citep{1999SoPh..188..3L}, length of the strong-field neutral line ($L_s$) \citep{2011SpWea...9.4003F} and mean photospheric excess energy density ($\rho_e$) \citep{1999A&AS..139..311B} of the source AR using HMI SHARP vector magnetogram just before the eruption. To calculate those parameters we consider pixels with vertical magnetic field intensity greater than $100$ G ($\mid B_z \mid > 100$ G) and the horizontal magnetic field strength greater than $200$ G ($\mid B_h\mid >200$ G) \citep[e.g.][]{2015GeoRL..42.5702T}. Thus, we remove the noisy pixels and those which do not belong to the AR. In Table 5, we give the measured values of the parameters. Using $\alpha_{AR}$, we calculate the overall global twist of the AR as $\alpha_{AR} L_{AR}$, where $L_{AR}$ is approximated as the length of the semi-circular field line with a radius of half of the distance between positive and negative flux weighted centers of the AR. We calculate $L_{AR}$= $7.8\times10^7$m thus $\alpha_{AR} L_{AR}$= 0.73. The overall twist of the MC ($\alpha_{mc} L_{mc}$) is calculated as 40 when $L_{mc}= 2$ AU, 50 when $L_{mc}= 2.5$ AU and 63 when $L_{mc}= \pi$ AU. For each of the $L_{mc}$s, the MC twists are an order of two greater than that of the AR. This result is consistent with what \citet{2004JGRA..109.5106L} found in their study of AR and MC overall twist with $L_{mc}$= 2.5 AU. It means the global twist of AR fails to estimate the twist invoked due to reconnection. The positive value of $\alpha_{AR}$ suggests that the rotation of magnetic field line in the flux rope at the source is right-handed which is consistent with the rotation of the magnetic field in 1-AU flux rope. It indicates that in this case the direction of magnetic field rotation in flux rope does not change after reconnection.   \par

\begin {table}
\begin{center}
\caption {Nonpotential magnetic parameters of AR 11692 }
\centering
 \begin{tabular}{ p{2cm} p{2cm} p{2cm} p{2cm}}\\
  \hline
  $I_z$&$\alpha_{AR}$&$L_s$&$\rho_e$\\
  $(10^{13}$ A)&(Mm$^{-1}$) & $(10^{7}$ m) & $( 10^{21}$ erg/cm)  \\
  (1)&(2)&(3)&(4)\\

  \hline
  1.58 & +0.01&3.23&7.95\\
 \hline
\end{tabular}

\end{center}

\end{table}
\begin{figure}[!htb]
\centering
  \includegraphics[width=0.9\textwidth]{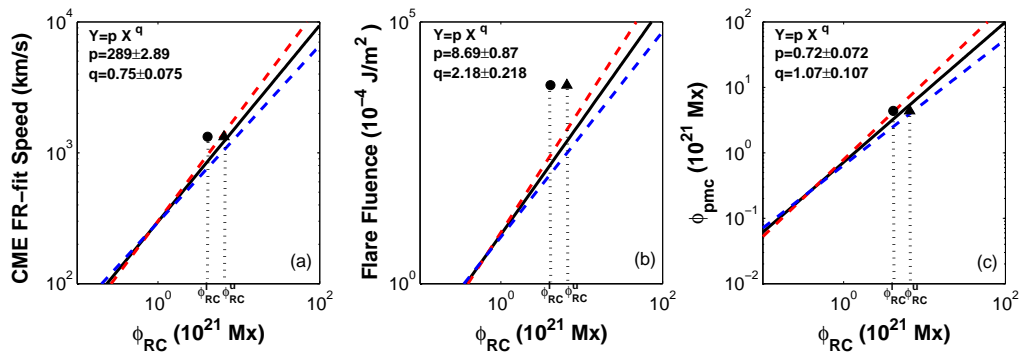}
   \caption{Linear regression models derived in \citet{2017barXiv170508912G} for the statistical relationship between $\phi_{RC}$ and (a) CME FR-fit speed, (b) Flare fluence, (c) $\phi_{pmc}$. The red and blue dotted lines are respectively upper(+10\%) and lower (-10\%) uncertainty levels of the model equations. The circular and triangular symbols corresponds to $\phi_{RC}^l$ and $\phi_{RC}^u$.}
\end{figure}

We compare our results with those of \citet{2017barXiv170508912G}, who studied 23 MCs among 54 ICMEs of solar cycle 23, computed their near-Sun and 1-AU parameters and obtained the relationship between them. In each of the panels of Figure 7 we plot the linear fit curve (in black firm line) with 10\% uncertainty in coefficients (p and q) (in red and blue dashed lines) derived from the statistical relationship between $\phi_{RC}$ and CME flux rope fit (FR-fit) speed, flare fluence and poloidal flux of MC. The circular and triangular points plotted on the figure, corresponds to $\phi_{RC}^l$ and $\phi_{RC}^u$. Figure 7(a) and 7(c) show that the measured values of CME speed and $\phi_{pmc}$ corresponding to the $\phi_{RC}^l$ and $\phi_{RC}^u$ almost match the values estimated using the equations of linear fit curves, whereas Figure 7(b) shows that according to the linear fit curve equation, the estimated flare fluence corresponding to the measured $\phi_{RC}$ range should be around half of the observed value in our case. From the distribution of $\phi_{RC}$, $B_{0cme}^{10R_s}$, and $R_{0cme}^{10R_s}$ of 20 MC events (see Figure 5 of \citet{2017barXiv170508912G}) of solar cycle 23, we notice that the range of $\phi_{RC}$ and $R_{0cme}^{10R_s}$ is smaller than the average values (respectively, $8.8\times10^{21}$ Mx and 4.12 R$_s$ ) derived from the distribution, whereas the $B_{0cme}^{10R_s}$ is greater than the average value of the distribution i.e. 51.9 mG. Furthermore, at 1 AU the values of $\phi_{pmc}$, $B_{0mc}$ and $R_{0mc}$ is smaller than the average values found from the distribution of 17 MC events (see Figure 6 of \citet{2017barXiv170508912G}). \par

 \section{Summary and Conclusions}

The main purpose of our study is to compare the relative magnetic helicity of ICME flux ropes at 1 AU and their solar source with the aim of understanding the origin and evolution of helicity of interplanetary flux ropes. For this study, we select the 17-18 March 2013 ICME as it has a clear solar source and 1-AU information. We compute the helicity of the flux rope using its axial magnetic field strength and physical parameters such as radius of its cross section and total length. We find a correspondence within 83\% between the measured quantity of helicity at solar source and 1 AU when the ICME flux rope is estimated as a half torus with total length of $\pi$ AU.\par

The main conclusions are,
\begin{enumerate}
    \item The helicity of MC at the source and 1 AU is broadly consistent when MC length is $\pi$ AU or the statistical value is $2.6\pm0.3$ AU considering the flux rope is still attached to the Sun.
    
    \item The amplitude of $H_{cme}$ is greater than that of $H_{mc}$, which suggests that 1-AU ICME flux rope helicity is primarily invoked by low coronal reconnection at the time of the eruption.

\end{enumerate}

Extracting the helicity information of CME at the source and 1 AU is challenging because 1) flux rope fitting results of MC and CME is not always perfect for each event, 2) the solar wind data are only available at the localized position of the satellite and 3) a major uncertainty exists in estimating MC length as well as CME flux rope length. However, this study indicates that it is crucially important to perform more comprehensive analysis of large data bases of MC events together with their source information. This may eventually lead to better estimates and predictions of the magnetic properties of ICME flux ropes at 1 AU and help us ascertaining their geo-effectiveness in advance.

\acknowledgments

This research was performed at the NASA Goddard Space Flight Center under a aegis of a SCOSTEP Visiting Scholar program. We are thankful to SCOSTEP, Catholic University of America and the Center of Excellence in Space Sciences India at IISER Kolkata for their support in carrying out this work. CESSI is funded by the Ministry of Human Resource Development, Government of India. We are grateful to the World Data Center for Geomagnetism, Kyoto teams for providing the solar wind and Dst data. We acknowledge the use of data from SDO, STEREO, SOHO, ACE and GOES instruments.
 

\end{document}